\documentclass[12pt]{iopart}
\usepackage{graphicx}
\usepackage{amsbsy}
\usepackage{color}

\begin{document}
\title{Gamma Rays from Centaurus A}

\author{Nayantara Gupta\dag}

\address{\dag\ Department of Physics, Indian Institute of Technology Bombay, Powai, Mumbai 400 076, INDIA}

\eads{\mailto{nayan@phy.iitb.ac.in}}

\begin{abstract}
Centaurus A, the cosmic ray accelerator a few Mpc away from us is possibly one of the nearest sources of extremely high energy cosmic rays. We investigate whether the gamma ray data currently available from Centaurus A in the GeV-TeV energy band can be explained with only proton proton interactions. 
We show that for a single power law proton spectrum, mechanisms of $\gamma$-ray production other than proton proton interactions are needed inside this radio-galaxy to explain the gamma ray flux observed by EGRET, upper limits by H.E.S.S./CANGAROO-III and the correlated extremely energetic cosmic ray events observed by the Pierre Auger experiment. In future with better $\gamma$-ray data, simultaneous observation with $\gamma$-ray and cosmic ray detectors, it would be possible to carry out such studies on different sources in more detail.  
 \end{abstract} 
keywords: ultra high energy cosmic rays, ultra high energy photons and neutrinos   
\maketitle
\section{Introduction}
Centaurus A (from now on Cen A) has drawn much attention of the cosmic ray and astrophysics community for its proximity and specially after the data from Pierre Auger (PA) \cite{auger1} experiment showed the correlation of atleast two events above 60 EeV with this radio-galaxy. 
From their data at the most ten events can be correlated with this source \cite{auger2}.  Although, the results from PA experiment do not identify Cen A as a source of extremely energetic cosmic rays, they suggest some of the events may 
come from this source. Being at a distance of 3.4 Mpc this radio loud galaxy has been observed in the radio to $\gamma$ ray wavelengths. 
The first results of high energy photon detection from Cen A were reported by Grindlay et al. \cite{grind}. Later Buckland Park array \cite{clay} and the JANSOZ observatory \cite{allen1,allen2} measured the gamma ray flux near 100TeV from this source. CANGAROO-I \cite{rowell}, JANZOS \cite{allen1,allen2} and Durham \cite{dur} observed the nuclear region of Cen A and set upper limits on the emission of very high energy photons from this region. H.E.S.S. group \cite{aha1} observed Cen A for a few hours in 2004 and put upper limit on gamma ray flux above 190 GeV. CANGAROO-III \cite{kabu} observed this source for approximately 10hrs in 2004 and obtained upper limits for various energy thresholds and different regions of the source.      
More photon data in the $\gamma$-ray wavelength are needed to understand the possible mechanisms of high energy photon production inside this source. Neutrino event rates have been calculated from Cen A \cite{cuo,hal} for future  kilometer scale neutrino detectors using the cosmic ray and $\gamma$-ray data. Inside astrophysical sources of cosmic rays, shock accelerated cosmic ray nuclei are expected to lose energy in proton proton ($pp$), proton photon ($p\gamma$) interactions, photo-disintegration and by synchrotron radiation of protons. Shock accelerated relativistic electrons cool by synchrotron radiation and inverse Compton scattering by low energy photons. Inside different sources different mechanisms dominate depending on whether most of the energies are carried by the electrons, the magnetic fields or the protons. $pp$ interactions are expected if matter density is high inside the source. $\pi^0$, $\pi^+$ and $\pi^-$ are produced with equal probabilities in $pp$ interactions. Subsequently $\pi^0$s decay to very high energy gamma rays and the charged pions produce very energetic neutrinos. If low energy photon density is high then one expects $p\gamma$ interactions. $\pi^0$, $\pi^+$ are produced with 2/3 and 1/3 probabilities in this case. In the final state one observes high energy gamma rays and neutrinos. Inverse Compton scattering of energetic electrons may also lead to the production of high energy photons. Low energy photons are expected to be generated in synchrotron radiation by relativistic electrons. The high energy photons are expected to interact with the low energy photons before escaping from a source. The $\gamma$-ray flux from a source is attenuated in this way. TeV photons are not significantly absorbed by the intergalactic infrared background if the source is located very close to us. A large number of AGN (Active Galactic Nuclei) have been observed in TeV $\gamma$-rays which helped in modeling the high energy photon absorption by the infrared radiation background \cite{stecker1}. At higher energies the 
 attenuation of gamma rays by cosmic microwave and infrared background is important even for sources very close to us. For detailed discussions one may see \cite{stecker2} 

$\gamma$-ray flux from a source is often used to estimate the neutrino event rate from it. This can be done if the $\gamma$-rays and neutrinos are produced together in similar kind of interactions ($pp$ or $p\gamma$). However, for a large number of sources it may not be the case. Various other processes may be involved in the production of $\gamma$-rays as discussed before. If leptonic processes are involved in the production of $\gamma$-rays then the estimation of neutrino flux using the $\gamma$-ray flux is expected to give incorrect result. In this context it is important to study the cosmic ray accelerators close to us.  
We investigate whether the photon data currently available from Cen A in the GeV-TeV energy range can be explained as only due to $pp$ interactions.  
\section{Correlating $\gamma$ Ray and Cosmic Ray Emission}
OSSE, COMPTEL and EGRET on board the Compton Gamma Ray Observatory (CGRO) observed Cen A. OSSE \cite{osse} and COMPTEL \cite{comp} data at keV and MeV energies yields a smooth, continuous spectrum that appears to evolve from a power law  above 200 keV with spectral index 1.97 and steepen gradually above 1MeV. This source has an one-sided X-ray jet collimated in the direction of its giant radio lobes. In the energy range of 30 to 10000 MeV EGRET observed Cen A at a confidence limit of $6.5\sigma$ \cite{sree}. EGRET localization and spectral measurements provide unique confirmation that the source detected by OSSE and COMPTEL is Cen A. This source provided the first evidence for $>100$ MeV emission with a confirmed large-inclination jet. The average flux above 100 MeV was found to be $(13.6\pm 2.5)\times 10^{-8}$ photons $cm^{-2} sec^{-1}$. The photon spectrum was fitted with a single power law of index $2.40\pm 0.28$. 
This is steeper than the average power law spectrum from $\gamma$-ray blazars (spectral index $2.15\pm 0.04$) and also steeper than the observed extragalactic  $\gamma$-ray background (spectral index $2.10\pm 0.03$).
The low $\gamma$-ray luminosity of Cen A is $L_{\gamma_l}\sim 10^{41}$ergs/sec. Typical $\gamma$-ray blazars are $10^5$ times more luminous than Cen A. 
Although Cen A was not observed by EGRET and PA experiment simultaneouly, our current analysis is based on the assumption that the MeV-GeV $\gamma$-ray flux from Cen A was similar to that observed by EGRET during the period of observation by PA experiment. In future if this source is observed simultaneously by the gamma ray and cosmic ray detectors, the observational data would be more useful for  our study.

At TeV energies this source was observed by Grindlay et al. \cite{grind}. 
JANSOZ array reported a search for gamma ray emission from Cen A over the period of October 1987 to January 1992 \cite{allen1,allen2}. No evidence was found for a steady  flux but a period of 48 days in 1990 showed a flux excess of $3.8\sigma$. The average flux detected was $(5.5\pm 1.5)\times 10^{-12} cm^{-2} sec^{-1}$ above 110TeV.
Buckland Park air shower array also observed this source and found excess of events with a confidence level of $99.4\%$ \cite{clay}. The average flux over the period of July 1984 to May 1989 was observed to be $(7.4\pm 2.6 )\times 10^{-12} cm^{-2} sec^{-1}$ between $\gamma$-ray energies 100 and 200TeV. An interpretation compatible with both experiments was given  as possibly Buckland Park array observed emission mainly prior to October 1987.
 It might be during the period of July 1984 to May 1985. Then Cen A remained in a low state until the JANSOZ group observed a flare in 1990. The average high state flux during 1984 to 1985 would then have been $2.9 \times 10^{-11} cm^{-2} sec^{-1}$.  The photon data near 100TeV energy has to be corrected for attenuation by background radiations (cosmic microwave and infrared) to find the intrinsic photon flux. The gamma ray observations carried out during the observations by PA experiments were by H.E.S.S. \cite{aha1} and CANGAROO-III \cite{kabu}. The upper limit on the integral photon flux assuming a power law of spectral index 3 has been obtained to be $5.68\times10^{-12} cm^{-2} sec^{-1}$ above energy 190 GeV by H.E.S.S. If extrapolated to 300 GeV this upper limit is one order of magnitude lower than the flux measured by Grindlay et al. \cite{grind}. Cen A was most likely in a lower emission state during observation by H.E.S.S. During a similar low emission state EGRET detected $>100$ MeV $\gamma$-ray emission from Cen A. If extrapolated to 190 GeV the EGRET spectrum yields a $\gamma$-ray flux which is $60\%$ of the upper limit on $\gamma$-ray flux obtained by H.E.S.S. The H.E.S.S. upper limit is $5.5\times10^{-12} cm^{-2} sec^{-1}$ when the spectral index is assumed to be $2.40$ similar to the spectral index used to fit EGRET data.
The upper limits on $\gamma$-ray flux reported by CANGAROO-III \cite{kabu} are comparable to the results reported by H.E.S.S. \cite{aha1}.

Cen A showed variability not only in $\gamma$-rays but also in X-ray and radio wave lengths. The possible mechanisms of $\gamma$-ray production inside astrophysical objects are inverse Compton scattering of electrons, proton-photon and proton-proton interactions. Also $\gamma$-rays can be produced in photo-disintegration and photo-de-excitaion of nuclei \cite{anchor}. As extreme energy cosmic ray events have been correlated with Cen A, photo-disintegration is expected to be less likely inside this source. Lower energy photons are expected from electron synchrotron radiation. Multi-band photon radiation of the knot in Cen A has been seen to be consistent with a magnetic field of strength $10^{-3}G$ and electrons' energy between 0.5 GeV and 5 GeV \cite{mao} accelerated by diffusive shock accelerations. Cen A jet has also been studied recently by Hardcastle et al. \cite{hard} in X-ray wavelength with $\it {Chandra}$. They have suggested that an unknown distributed particle acceleration process operates in the jet of Cen A. We have even less knowledge about the nucleus of Cen A.  Infrared imaging polarimetry of the nuclear regions of Cen A revealed a very red polarized source which was interpreted as a blazar type nucleus \cite{bailey}. The mid-IR images of the nuclear region of Cen A revealed a linear size of approximately 3 pc \cite{karovska,radomski}. In the nuclear region matter and photon densities are expected to be high. If TeV energy $\gamma$-rays can escape from the nuclear region then the mean free path of $\gamma \gamma$ interactions has to be larger than the size of the source. On the other hand it may also be possible that the TeV photons are mostly absorbed inside the source. In that case the observed photon flux is much smaller than the one produced inside the source. This issue would be resolved in future with improved data, if $\gamma$-ray detectors can get the photon spectrum at TeV energies from Cen A. 
The observed $\gamma$-ray spectrum would show a sharp cut-off if there is significant absorption inside the source. The cross-section of interactions of protons with low energy photons is more than three 
orders of magnitude smaller than the cross-section of pair production \cite{ner}. The cross-sections of $pp$ interactions are much higher than $p\gamma$ interactions \cite{pd}. The shock accelerated protons can interact by both $p\gamma$ and $pp$ interactions to produce gamma rays and neutrinos. Protons also lose energy by synchrotron radiation. The relativistic proton spectrum can be expressed as a power law.
\begin{equation}
\frac{dN_p(E_p)}{dE_p}=A_p E_p^{-\alpha}
\label{p_flux}
\end{equation}
A break is expected in the spectrum if the protons are losing energy in $p\gamma$ interactions. $p\gamma$ interactions take place when the resonance condition is satisfied $E_pE_{\gamma}\geq 0.3 GeV^2$ where $E_{\gamma}$ is the energy of low energy photons. In this  case the proton spectrum would be steeper above the break energy. A break in the shock accelerated proton spectrum can also arise due to strongly modified shocks if the approximation of test particles is relaxed \cite{alo}. However, in this case the proton spectrum will be flatter above the break energy unlike the case of $p\gamma$ interactions. In case of acceleration and subsequent reacceleration the proton spectrum may also become steeper at higher energies. From the current data it is not possible to say whether there is a break in the  intrinsic proton spectrum near the energy (60 EeV) of our interest. In $pp$ interactions $\pi^0$s are produced with a probability of $1/3$. Each of the two energetic photons produced through $\pi^0$ decay carries nearly $10\%$ of the initial proton's energy. We can express the ratio of the luminosity observed by EGRET \cite{sree} and the upper limit from H.E.S.S.  \cite{aha1} as follows assuming the photons are of nucleonic origin 
\begin{equation}
\frac{L_{\gamma,100MeV-10GeV}}{L_{\gamma,>0.19TeV}}=\frac{0.1^{2-\alpha}-10^{2-\alpha}}{ 190^{2-\alpha}}=\frac{10^{41}erg/sec}{2\times10^{39}erg/sec}=50 
\end{equation} 
We find $\alpha=2.5$ and $A_p\simeq 10^{-11} TeV^{-1}cm^{-2}sec^{-1}$ assuming the TeV photons are produced in $pp$ interactions inside Cen A at a distance of 3.4 Mpc. The photon luminosities inside the source are assumed to be same as the observed luminosities as we have used only the luminosities in the GeV-TeV energy range. If in future H.E.S.S./CANGAROO-III observes a lower flux of TeV  
$\gamma$-rays, using that we will get a higher value of $\alpha$.  
Grindlay et al. \cite{grind} reported positive detection of TeV $\gamma$-rays from Cen A. But we have not used their results here as most likely those observations  were carried out during a high state of Cen A.
The mean free path of $pp$ interactions is $\lambda_{pp}=\frac{1}{\sigma_{pp}n_p}$ where $\sigma_{pp}$ is the cross-section of $pp$ interactions and $n_p$ is the nucleon density. The optical depth of $pp$ interactions $\tau_{pp}=D/\lambda_{pp}$ where $D$ is the size of the nuclear region. Moreover, for every target proton there  is an electron contributing to the Thomson optical depth. If the source is optically thin $\tau_{pp}$ is constrained by the ratio of $pp$ and Thomson scattering cross sections. $\tau_{pp}$ can not be more than 0.03 for optically thin sources. The number of cosmic ray events observed by Pierre Auger observatory can be expressed as
\begin{equation}
N_{obs}(E_p>60EeV)=A_d  \omega(\delta) E_p \frac{dN_p(E_p)}{dE_p} exp(-\tau_{pp})\times 3.15\times10^7 
\label{cr_ev}
\end{equation}
The integrated exposure of PA for a point source is $A_d=\frac{9000}{\pi} km^2 yr$, $\pi$ sr is the PA field of view corresponding to 60 degrees as maximum zenith angle. The data was collected during a period of 15/4 years between $1^{st}$ January 2004 and August 2007. The factor $\omega(\delta)$ is relative exposure for angle of declination $\delta$. Cen A is at $\delta=-47^{\circ}$ and the corresponding value for $\omega(\delta)$ is about 0.64 \cite{cuo}.
Two events have been correlated with Cen A above $60$EeV.  
 If we use the proton spectral index $\alpha=2.5$ and $A_p=10^{-11} TeV^{-1} cm^{-2} sec^{-1}$ we find $exp(\tau_{pp})<1$ from eq(\ref{cr_ev}) at $E_p=60 EeV$, which is not physically possible. Thus for a single power law proton spectrum, $pp$ origin of photons in the entire GeV to TeV energy range seems to be inconsistent with the extreme energy cosmic ray data. The general condition for ruling out the $pp$ origin of $\gamma$-rays is $A_d \omega(\delta) A_p/(N_{obs}(E_p>60EeV)E_p^{(\alpha-1)})\times3.15\times10^7<1$ at $E_p=60$ EeV. Future data would help us to determine the production mechanisms of $\gamma$-rays more precisely.    

 \section{Discussion and Conclusion}
 We have used the $\gamma$-ray flux from EGRET \cite{sree} and the upper limit on TeV $\gamma$-ray flux from H.E.S.S. \cite{aha1} in our current analysis. 
H.E.S.S. and CANGAROO-III observed Cen A when PA was also collecting data. Assuming a single power law spectrum of shock accelerated protons all the way from GeV to EeV energy, the $pp$ origin of $\gamma$-rays at both GeV and TeV energies appears to be inconsistent with the current observational data. 
Other mechanisms of $\gamma$-ray production should be considered in modeling this source. The low energy photon density has to be high for significant yield of high energy photons by inverse Compton scattering of electrons and $p\gamma$ interactions. In the case of $p\gamma$ interactions a break is expected in the proton spectrum and as a result in the $\gamma$-ray, neutrino spectra. 
Also, significant absorption of high energy photons inside the source is expected if the low energy photon density is high. 
As discussed earlier a break in the proton spectrum may also appear due to shock
 accelerations.
Due to insufficient $\gamma$-ray and cosmic ray data, currently it is not possible to understand the underlying mechanism of $\gamma$-ray production in Cen A. The present \cite{hess,milagro,ver,agile} and upcoming \cite{glast} $\gamma$-ray, cosmic ray detectors should observe Cen A and help us to explore the physical processes going on inside the sources of extremely energetic cosmic rays.
\section{Acknowledgement}
The author thanks the anonymous referee for important comments which helped to improve this paper and also to Bing Zhang for helpful communications. 

\end{document}